\documentclass[conference]{IEEEtran}
\IEEEoverridecommandlockouts
\usepackage{cite}
\usepackage{amsmath,amssymb,amsfonts}
\usepackage{algorithmic}
\usepackage{graphicx}
\usepackage{textcomp}
\usepackage{xcolor}
\usepackage{url}
\usepackage{hyperref}

\def\BibTeX{{\rm B\kern-.05em{\sc i\kern-.025em b}\kern-.08em
    T\kern-.1667em\lower.7ex\hbox{E}\kern-.125emX}}

\begin{document}

\title{Exploring Non-Linear Programming Formulations in QuantumCircuitOpt for Optimal Circuit Design}

\author{
\IEEEauthorblockN{Elena R. Henderson}
\IEEEauthorblockA{\textit{Department of Computer Science} \\
\textit{Southern Methodist University}\\
Dallas, TX, USA \\ 
erhenderson@smu.edu}
\and
\IEEEauthorblockN{Harsha Nagarajan}
\IEEEauthorblockA{\textit{Applied Mathematics and Plasma Physics (T-5)} \\
\textit{Los Alamos National Laboratory}\\
Los Alamos, NM, USA \\
harsha@lanl.gov}
\and
\IEEEauthorblockN{Carleton Coffrin}
\IEEEauthorblockA{\textit{Advanced Network Science Initiative} \\
\textit{Los Alamos National Laboratory}\\
Los Alamos, NM, USA \\ 
cjc@lanl.gov}
}

\maketitle

\begin{abstract}
Given the limitations of current hardware, the theoretical gains promised by quantum computing remain unrealized across practical applications. But the gap between theory and hardware is closing, assisted by developments in quantum algorithmic modeling. One such recent development is QuantumCircuitOpt (QCOpt), an open-source software framework that leverages state-of-the-art optimization-based solvers to find provably optimal compact circuit decompositions, which are exact up to global phase and machine precision. The quantum circuit design problem can be modeled using non-linear, non-convex constraints. However, QCOpt reformulates these non-linear constraints using well-known linearization techniques such that the resulting design problem is  solved as a Mixed-Integer Linear Programming (MILP) model. In this work, we instead explore whether the QCOpt could also be effective with a continuous Non-Linear Programming (NLP) model obtained via relaxation of the integer variables in the non-linear constraints. We are able to present not only multiple significant enhancements to QCOpt, with up to \textit{11.3x} speed-up in run times on average, but also opportunities for more generally exploring the behavior of gradient-based NLP solvers. 
\end{abstract}


\section{Introduction}
\label{sec:intro}
Quantum computing promises to alter the landscape of our computer-driven world, improving everything from finance to medicine to national defense \cite{Egger_finance, Flother_healthcare, QIS_national, Preskill_quantum}. 
These gains remain only theoretical, however, given the physical limitations of today’s quantum computers \cite{Preskill_quantum}. 
But the gap between theory and hardware is closing, assisted by development in quantum algorithmic modeling \cite{Gambetta_ibm, Vatan_optimal}.

Quantum algorithms are typically formalized as quantum circuits, which describe any algorithm as an ordered sequence of quantum gates that act on quantum bits, or ``qubits." Though a plethora of quantum gates exist in theory, current quantum computers physically implement only a limited set of quantum gates. 
Every quantum computer has a specific so-called ``native gate set".  
Thus, in order to run a given quantum algorithm on a given quantum computer, the algorithm must be represented using a limited set of native gates. 
It is well-known that in theory, any $n$-qubit quantum algorithm that is represented as a ``target gate'' can also be represented as a sequence of one- and two-qubit gates \cite{barenco1995elementary}. However, in practice, the circuit representation is constrained by hardware limitations if it is to successfully run to completion on current quantum machines. 

\begin{figure}[h]
\centerline{\includegraphics[scale=0.16]{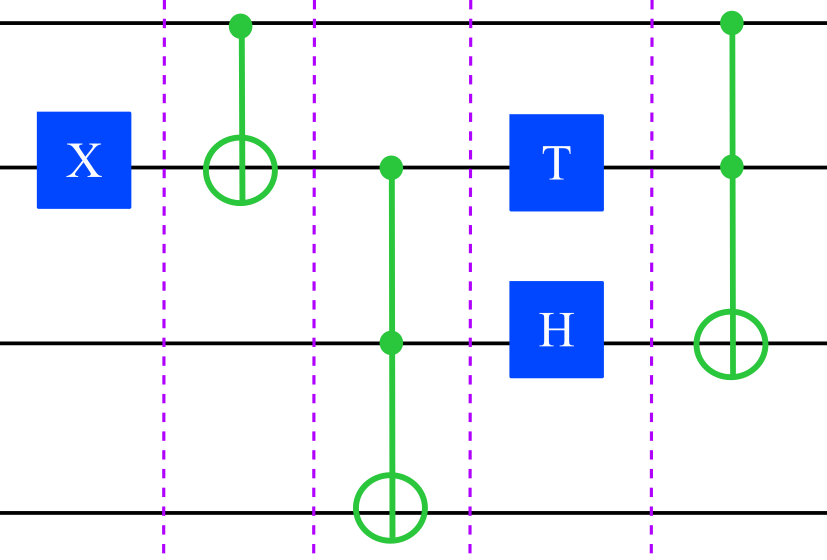}}
\caption{A quantum circuit with a depth of five, as illustrated by dashed lines. The second qubit (black line) from the top is operated on by five gates, the most of any qubit in the circuit.}
\label{fig:depthfigure}
\end{figure}

Two of the most critical hardware limitations of today's quantum computers are (1) circuit ``depth"--here measured as the longest gate path on a single qubit--and (2) the number of entangling, two-qubit CNOT gates in a circuit. 
(Figure \ref{fig:depthfigure} provides an illustration of the circuit depth.)
The shorter the depth of a circuit and the fewer CNOT gates it contains, the more likely the circuit will run to completion, as it is less susceptible to both quantum noise and quantum decoherence, the latter of which renders a circuit's qubits inoperable \cite{Shor_decoherence}. However, given these limitations, even for two-qubit circuits, finding such efficient circuit implementations with theoretical guarantees can be forbiddingly complex \cite{Vatan_optimal}. 

In order to address these limitations, there exist several commercial and open-source software tools for circuit synthesis, such as IBM's Qiskit \cite{ibm_qiskit}, Google's Cirq \cite{Cirq_website}, BQiskit \cite{davis2020towards} and QuantumCircuitOpt \cite{Nagarajan_qcopt}. 
The latter, hereinafter referred to as QCOpt, is an open-source framework \footnote{\url{https://github.com/harshangrjn/QuantumCircuitOpt.jl}} that leverages state-of-the-art optimization-based solvers to find provably optimal compact circuit decompositions \cite{Nagarajan_qcopt}. Specifically, given a set of one- and two-qubit gates--which is often the native gate set of a certain quantum computer--and a target gate, QCOpt finds a circuit decomposition with the shortest possible depth and/or with the fewest possible CNOT gates \cite{Nagarajan_qcopt}.

The optimal quantum circuit decomposition problem can be formulated with a linear objective function and a set of non-linear, non-convex constraints, resulting in a Mixed-Integer Non-Linear Program (MINLP). QCOpt solves the MINLP by applying linearization techniques from optimization theory, reformulating the problem into an equivalent  Mixed-Integer Linear Programming (MILP) model \cite{Nagarajan_qcopt}. Although this MILP model was found to be effective, it is well-known that MILP solvers can be expensive \cite{gurobi}, particularly on larger qubit circuits. In this work, in lieu of explicit linearization techniques, we explore the efficacy of continuous non-linear programming (NLP) models within QCOpt, obtained from the relaxation of integer variables in the aforementioned MINLP model. Our study into this model has produced multiple enhancements for QCOpt, as well as opportunities for exploring efficient large-scale barrier methods developed for NLPs on the order of millions of variables \cite{Wachter_ipopt, byrd2006k}.

This paper is organized as follows: In section \ref{sec:qcoptmodel}, we review the mathematical formulation/model within QCOpt and its variants, including the continuous NLP model that is the focus of this paper. 
Section \ref{sec:NLPBehavior} discusses the unexpected behavior of the NLP model that led to the numerical experiments in section \ref{sec:results} comparing the NLP model's performance with that of the MILP model. Finally, section \ref{sec:conclusions} concludes the paper with discussions on potential avenues for future research. 

\section{Mathematical models in QuantumCircuitOpt}
\label{sec:qcoptmodel}

\begin{figure}[h]
    \centering
    \includegraphics[scale=0.15]{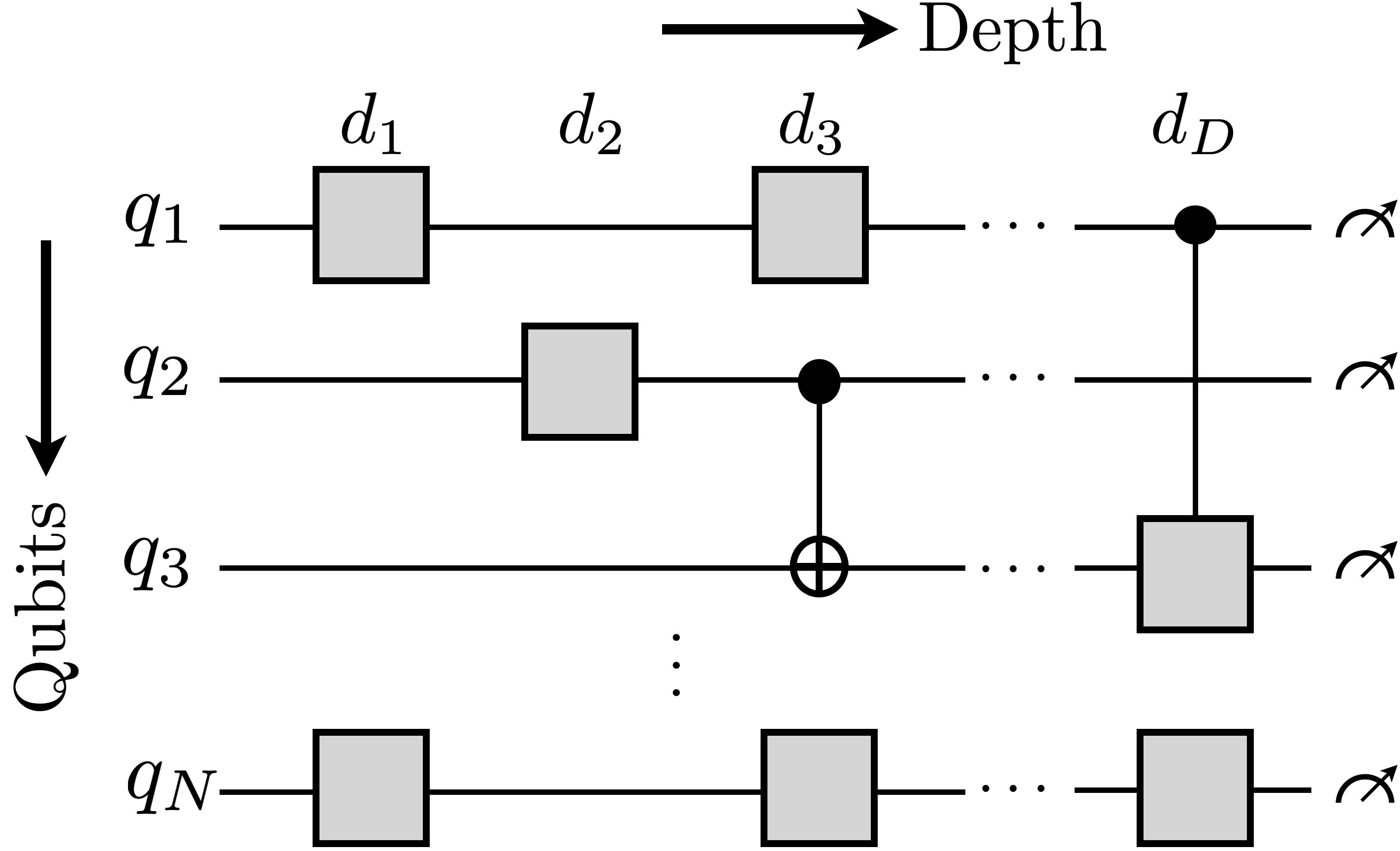}
    \caption{A generic quantum circuit of $N$ qubits and depth $D$, consisting of an ordered sequence of one- and two-qubit quantum gates.}
    \label{fig:circuit}
\end{figure}

In this section, we review the mathematical models (formulations) within the QCOpt package. 
As alluded to before, QCOpt produces an optimized quantum circuit that is composed of an ordered sequence of one- and two-qubit gates, such as that shown in Figure \ref{fig:circuit}. 

We now explain how QCOpt mathematically models such a circuit. For an arbitrary decomposed circuit, let $N$ be its number of qubits, let $D$ be its (user-defined) maximum allowable depth, and let $\mathrm{T^g}$ be the matrix representation of the target gate that this circuit is to represent. 
Additionally, let $\{\mathrm{M}_1, \mathrm{M}_2,\ldots, \mathrm{M}_K\}$ be the set of complex-valued $\left(2^N \times 2^N\right)$ matrices that represent the $K$ ``input gates'' (or the native gates) available for the decomposition. 

Given these circuit components, the optimization model variables that select input gates for a decomposition are as follows. 
Let $[z_{1,d},\ldots,z_{K,d}] \in \{0,1\}^{K,D}$ be binary variables that are parametrized by the depth of the circuit. 
These are used to assign one input gate to each of the $d$ depths of a circuit: a $z_{k,d}$ value of $1$ represents a choice of input gate $\mathrm{M}_k$ at depth $d$, and a $z_{k,d}$ value of $0$ represents the absence of input gate $\mathrm{M}_k$ at that depth. 
To describe the constraints that limit the selection of input gate choice, let $\mathrm{G}_d$ be a matrix of continuous variables that represents the choice of one of the input gates at depth $d$, such that
\begin{subequations}
    \begin{align}
        & \mathrm{G}_d = \sum_{k=1}^K z_{k,d}\mathrm{M}_{k}, \quad \forall d = 1,\ldots, D, \label{eq:Gd_constraint}  \\ 
        & \sum_{k=1}^K z_{k,d} = 1,\quad \forall d = 1,\ldots, D. \label{eq:single_gate}
    \end{align}
\end{subequations}
Unlike the standard formalization of quantum circuit gates, without any loss of generality, the QCOpt model assumes that each input gate must be located at its own separate circuit depth, that is, a single input gate must be selected for each circuit depth, as modeled in \eqref{eq:single_gate}. 

For example, in a two qubit circuit ($N=2$), let the native gates set include the universal $\mathrm{U_3}(\hat{\theta}, \hat{\phi}, \hat{\lambda})$ gate (which could be located on either of the qubits) and an entangling $\mathrm{CNOT}$ gate. Here $(\hat{\theta}, \hat{\phi}, \hat{\lambda})$ are the three discretized Euler angles parametrizing the $\mathrm{U_3}$ gate. Let `$\mathrm{I}$' be the one-qubit identity gate which is basis independent and does not modify the quantum state on the qubit in which it is located. Their matrix representations are: 
\begin{subequations}
\begin{align}
\mathrm{U_3}(\hat{\theta}, \hat{\phi}, \hat{\lambda}) & =
    \begin{bmatrix}
        \cos(\hat{\theta})          & -e^{\mathrm{i} \hat{\lambda}}\sin(\hat{\theta}) \\
        e^{\mathrm{i}\hat{\phi}}\sin(\hat{\theta}) & e^{\mathrm{i}(\hat{\phi}+\hat{\lambda})}\cos(\hat{\theta})
    \end{bmatrix},  \\
\mathrm{CNOT} &= 
\begin{bmatrix}
1 & 0 & 0 & 0 \\
0 & 1 & 0 & 0 \\ 
0 & 0 & 0 & 1 \\
0 & 0 & 1 & 0
\end{bmatrix}, \\ 
\mathrm{I} &= 
\begin{bmatrix}
1 & 0 \\
0 & 1
\end{bmatrix}. \label{eq:cnot_gate}
\end{align}
\label{eq:native_gates}
\end{subequations}
Then, the $\mathrm{G}_d$ matrix constraint in \eqref{eq:Gd_constraint} reduces to 
\begin{align}
& \mathrm{G}_d =  z_{1,d}\left(\mathrm{U_3}(\hat{\theta}, \hat{\phi}, \hat{\lambda}) \otimes \mathrm{I}\right) +  z_{2,d}\left( \mathrm{I} \otimes \mathrm{U_3}(\hat{\theta}, \hat{\phi}, \hat{\lambda})\right) \nonumber \\ 
& \hspace{1cm} + z_{3,d}(\mathrm{CNOT}) + z_{4,d}\left(\mathrm{I}^{\otimes 2}\right)  \quad \forall d=1,\ldots, D,
\end{align}
where $\otimes$ represents the Kronecker product of gates. \\
Finally, the multiplicative property of unitary gates, $\mathrm{G}_d$, on a quantum circuit which is intended to represent any target computation, $\mathrm{T^g}$, can be modeled as the following constraint:
\begin{align}
    \mathrm{G}_0 \prod_{d=1}^{D} \mathrm{G}_d = \mathrm{T^g}
    \label{eq:Gd_prod}
\end{align}
where $\mathrm{G}_0$ is the initial state of the circuit. 
Because this constraint of variable-matrix products cannot be implemented directly due to the limitations of state-of-the-art optimization solvers, we reformulate it as recursive bi-matrix product constraints as follows: 
\begin{subequations}
\begin{align}
    & \widehat{\mathrm{G}}_d = \widehat{\mathrm{G}}_{d-1} \mathrm{G}_d \quad \forall d=2,\ldots,(D-1), \label{eq:recursive}\\ 
    & \widehat{\mathrm{G}}_1 = \mathrm{G}_0 \mathrm{G}_1, \\ 
    & \widehat{\mathrm{G}}_{d-1}\mathrm{G}_D = \mathrm{T^g},
\end{align}
\label{eq:recursive}
\end{subequations}
\noindent
where $\widehat{\mathrm{G}}_d$ represents cumulative products of unitary gates, thus preserving the property of being a unitary matrix. The elements of this matrix can be values that lie within the range $[-1,1]$.  \\ 

\noindent
\textbf{Objective function} \\ 
To minimize the total depth of the circuit as an objective function for the optimization model, we minimize the number of one- and two-qubit gates (excluding the identity gate) admitted in the decomposition. This objective, which also serves as a proxy for the execution time of the circuit, can be modeled as a linear function as follows:
\begin{align}
  \mathrm{minimize} \quad  \left(\sum_{k \in \{1,\ldots, K | \atop \mathrm{M}_k \neq \mathrm{I}^{\otimes N} \}} \  \sum_{d=1}^D z_{k,d}\right) \label{eq:objective}  
\end{align}
The above objective function could also be easily generalized to other types of objectives like minimizing the total number of CNOT gates in the decomposable circuit \cite{Nagarajan_qcopt} or minimizing the cross-talk noise on quantum processors for the resulting circuit \cite{murali2020software}. However, in this paper, we will only focus on the depth minimization objective (as in \eqref{eq:objective}). \\

To summarize, the problem of optimal circuit design can be modeled as an MINLP that includes the above linear objective function and a set of bilinear, non-convex constraints \eqref{eq:recursive}.

\subsection{MILP model using linearizations}
\label{subsec:milp_model}
In \eqref{eq:recursive}, note that the elements of the $\widehat{\mathrm{G}}_d$ matrix represent bilinear terms which are the product of a continuous and a binary variable. Let one such term be a product of $g \in [-1,1]$ (any element of $\widehat{\mathrm{G}}_d$)  and $z \in \{0,1\}$ which represents a binary variable associated with one of the native gates. This product can be exactly reformulated by applying the following four linear constraints with an auxiliary variable $\widehat{gz}$ per product term as follows:
\begin{subequations}
\begin{align}
   \widehat{gz} \geqslant -z, \quad & \widehat{gz} \geqslant g+z-1, \\ 
   \widehat{gz} \leqslant z, \quad & \widehat{gz} \leqslant g-z+1.
\end{align}
\label{eq:mcc}
\end{subequations}
QCOpt implements an MILP model with  McCormick linearization constraints in \eqref{eq:mcc} \cite{mccormick1976computability, Nagarajan_qcopt}. Note that this MILP model's optimal solution is same as that of an MINLP model. Hence, given an efficient MILP solver, QCOpt provides optimal circuit decompositions for any target as a function of the input set of gates. 

Methods for solving MILPs, while NP-Hard in the worst case, have made dramatic strides in practical applications via state-of-the-art commercial solvers such as CPLEX and Gurobi \cite{gurobi}. MILP solvers have also been found very effective in benchmarking novel computational platforms like an adiabatic quantum computer \cite{coffrin2019evaluating}. 

\subsection{Enhancements via symmetry-breaking constraints}
\label{subsec:symmetry}
\begin{figure}[htbp]
\centerline{\includegraphics[scale=0.23]{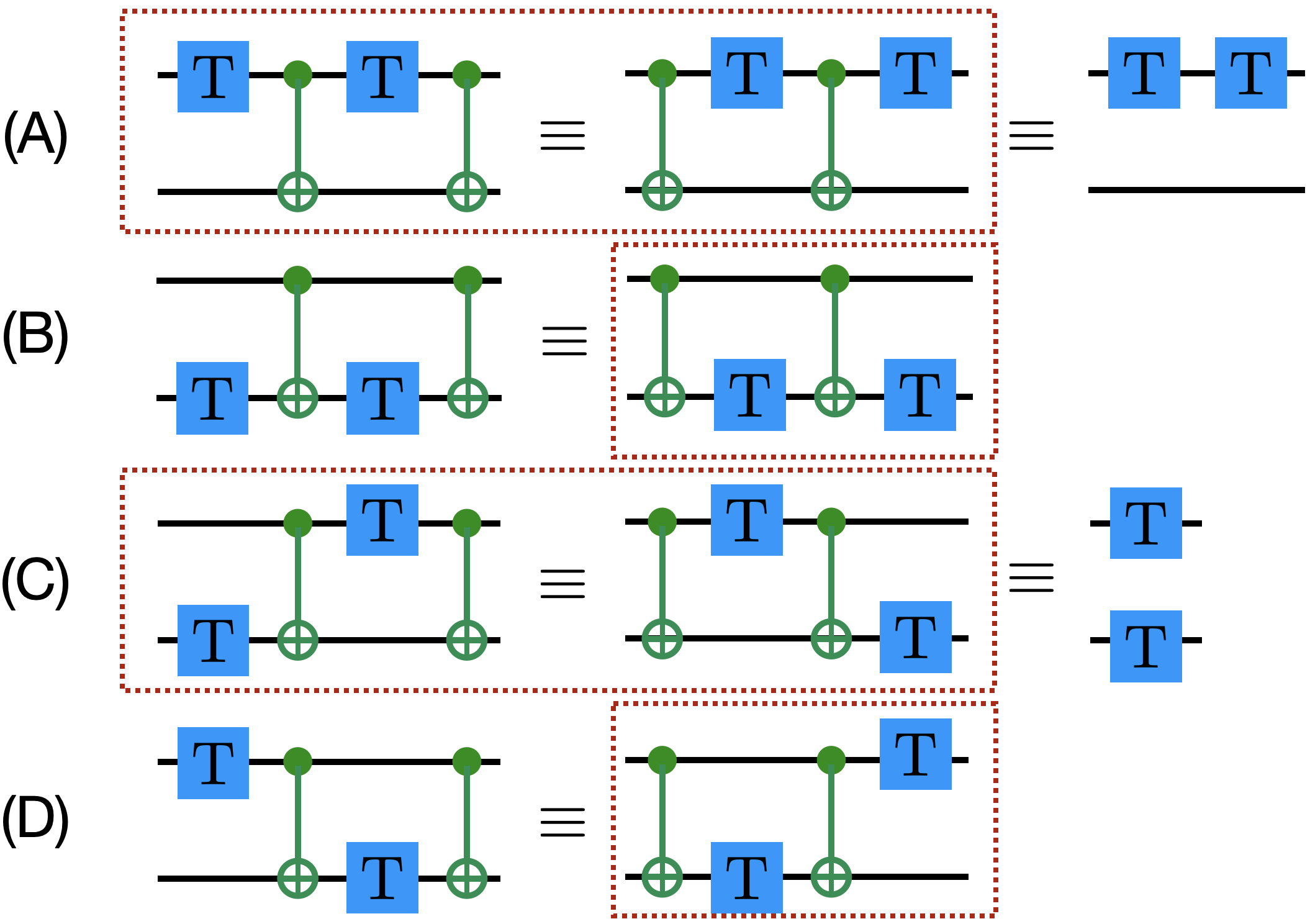}}
\caption{Four different gate sequences which are reverse-commutative are illustrated here. All the computationally equivalent gate sequences can be safely eliminated from the solution space by retaining one such feasible sequence. The eliminated gate sequences are highlighted within dotted boxes.}
\label{fig:sym_solutions}
\end{figure}
There are often symmetric solutions for QCOpt to represent an optimal circuit decomposition, since some gate sequences can be ordered in multiple ways while still producing the same effect on the quantum states of the corresponding qubits. This results in a prohibitively large solution space, which makes the problem more difficult to solve. 
So, QCOpt employs a set of ``symmetry-breaking'' valid constraints (linear functions of binary variables $z_{k,d}$) that prohibit the selection of redundant gate sequences in the search for an optimal sequence. For example, given an input gate set $\{\mathrm{T}_1, \mathrm{T}_2, \mathrm{CNOT}_{12}\}$, numerous redundant gate sequences are illustrated in Figure \ref{fig:sym_solutions}. 

In order to eliminate the gate sequences highlighted in dotted boxes in Figure \ref{fig:sym_solutions}, QCOpt adds the following set of linear, symmetry breaking, valid constraints for every depth $d \in 1,\ldots, (D-3)$: 
\begin{subequations}
\begin{align}
& z_{T_1, d} + z_{cnot, d+1} + z_{T_1, d+2} + z_{cnot, d+3} \leq 3, \\ 
& z_{cnot, d} + z_{T_1, d+1} + z_{cnot, d+2} + z_{T_1, d+3} \leq 3, \\ 
& z_{cnot, d} + z_{T_2, d+1} + z_{cnot, d+2} + z_{T_2, d+3} \leq 3, \\
& z_{T_2, d} + z_{cnot, d+1} + z_{T_1, d+2} + z_{cnot, d+3} \leq 3, \\
& z_{cnot, d} + z_{T_1, d+1} + z_{cnot, d+2} + z_{T_2, d+3} \leq 3, \\
& z_{cnot, d} + z_{T_2, d+1} + z_{cnot, d+2} + z_{T_1, d+3} \leq 3, 
\end{align}
\end{subequations}

\section{Non-linear Programming Relaxation Model}
\label{sec:NLPBehavior}

We now present the primary focus of this paper: the exploration of the behavior of a non-linear programming (NLP) model as a relaxation of the original MINLP model with bilinear, non-convex constraints derived in section \ref{sec:qcoptmodel}. A well-known approach to solve MINLPs of this type is to apply global optimization methods via non-linear branch-and-bound and piecewise polyhedral relaxation algorithms \cite{nagarajan2019adaptive, sahinidis1996baron, kroger2018juniper, nagarajan2016tightening}. However, since these methods are known to be computationally expensive, we apply a simple NLP-based relaxation which also serves as a primary initial step in the global optimization algorithms. More precisely, all the binary $z_{k,d}$ variables are relaxed, or made continuous: the solver can assign them to any real number in the range $[0,1]$, thus making this an NLP relaxation model of an MINLP. In this NLP model (or the NLP relaxation model), note that we do not explicitly apply any linearization of bilinear non-convex terms which appear in constraints \eqref{eq:recursive}. 
\vspace{0.35cm}

\noindent
\textbf{Solving the NLP model to global optimality} \\ 
The NLP relaxation of the MINLP model, although it is quadratic and non-convex, can be solved to compute its global optimum using the standard algorithms as described above. With an expanded solution space on $z_{k,d}$ variables, one would expect these continuous NLP models to be relatively weak, and thus choose fractional solutions for these variables. However, we observed numerically that the quality of this NLP model is very tight, and indeed produces only ``integral solutions'' as the global optimum. This observation was quite intriguing, although it was possible only with very small circuits when using Gurobi as the underlying non-convex global solver \cite{gurobi}. Although the globally optimal solutions of this NLP model and the MILP model in section \ref{subsec:milp_model} were identical, the run times of the MILP model were much faster, indicating the NLP model is a much more difficult problem. \\

\begin{figure}[ht]
    \centering
    \includegraphics[scale=0.87]{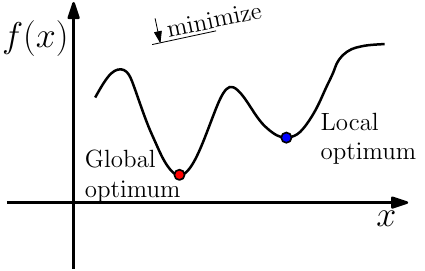}
    \caption{A simplified illustration of local vs. global optimal solutions while minimizing a one-dimensional function.}
    \label{fig:local_global}
\end{figure}

\noindent
\textbf{Solving the NLP model to local optimality} \\ 
Since solving NLP models to global optimality was very expensive and impractical, we wanted to see if we could obtain this seemingly powerful NLP model behavior on the decomposition problem of larger circuits. Hence, we instead solved the NLP model to local optimality using efficient gradient-based barrier methods in solvers such as Ipopt \cite{Wachter_ipopt} and Knitro \cite{byrd2006k}. Although these solvers guarantee a local solution which is feasible to the NLP, the solution may not necessarily be a global optimum (as shown in Figure \ref{fig:local_global}). \textit{However, we observed that any feasible solution to the NLP relaxation model is not only integral, but is also globally optimal in most of the cases, and that these solvers are able to find such solutions in run times no more than a few minutes.} This intriguing behavior of the proposed NLP model is rather unexpected and is promising for the exploration of theoretical analysis. 

Finally, we also implemented a linear programming (LP) model, a continuous version of the MILP model in section \ref{subsec:milp_model} with the only change in implementation being the relaxation of binary variables to values in $[0,1]$. The performance of the LP model was expectedly lackluster, with fractional values assigned to the $z_{k,d}$ variables, thus making it non-interesting from a quantum circuit designer's perspective. 

In the next section, we compare the performance of the NLP relaxation model with different solvers, as well as the performance of the MILP model from section \ref{subsec:milp_model}.  

\section{Case studies for Circuit Decompositions}
\label{sec:results}
In this section, we provide proof-of-concept case studies to demonstrate the efficacy of using proposed optimization models (NLP and MILP) within QuantumCircuitOpt (v0.5.0), implemented in Julia programming language \cite{bezanson2017julia,lubin2022jump}, for obtaining optimal quantum circuit designs. All computations were performed on an Intel Core i7 machine running at 2.30 GHz with 16 GB of RAM running on the Windows platform. 

\subsection{Decomposition Specifications}

Tables \ref{tab:gate_specs_const} and \ref{tab:gate_specs_param} describe the two, three, and four qubit representative benchmarks for circuit decompositions we use to asses the performance of the MILP vs. the NLP optimization model. The primary difference between these tables is that Table \ref{tab:gate_specs_const} represents the specifications for input native gates with constant parameters (such as gates from the Clifford + $\mathrm{T}$ group), while Table \ref{tab:gate_specs_param} represents the specifications for input gates with discretized rotation angle parameters for $\mathrm{U_3, R_x, R_y, R_z}$ gates. Also, the $N$-qubit `$\mathrm{Identity}$' gate ($\mathrm{I}^{\otimes N}$) and the two-qubit `$\mathrm{CNOT}$' gate are part of all the input gates in both of the tables. For example, for the 2-qubit target gate, `Reverse-CNOT' in Table \ref{tab:decomp_specs_const_caption}, the input gate set consists of the $\mathrm{H}_1, \mathrm{H}_2, \mathrm{CNOT}_{1,2}$ and $\mathrm{I}^{\otimes 2}$ gates. Given these gates, and user-defined parameters such as the maximum allowed depth of $5$ for the circuit to be decomposed, the output of QCOpt--after solving the underlying optimization model-- will be a circuit of optimal (best) depth equal to $3$. This optimal circuit is given by $(\mathrm{H}_1 \otimes \mathrm{H}_2)\cdot \mathrm{CNOT}_{1,2} \cdot (\mathrm{H}_1 \otimes \mathrm{H}_2)$. Similar descriptions hold true for all the other gates in both the tables. When angle-parametrized input gates are used in a decomposition (in Table \ref{tab:gate_specs_param}), QCOpt allows the user to provide a set of angle discretization values. For more details on this, we refer the reader to \cite{Nagarajan_qcopt}. Detailed input settings for every test target gate and input gate set in this section are available at this open-source link: \textit{\url{https://github.com/harshangrjn/QuantumCircuitOpt.jl/tree/master/examples}}. 

In both Tables \ref{tab:gate_specs_const} and \ref{tab:gate_specs_param}, the target gates are ordered by a rough measurement of decomposition difficulty, in order of increasing difficulty from top to bottom. This ``difficulty'' measure can be roughly attributed due to the following input parameters: the number of qubits ($N$), the number of input native gates and the maximum 
allowable depth of the circuit ($D$).

\begin{table}[ht]
\caption{Circuit decomposition specifications with Clifford+$\mathrm{T}$ gates.}
\label{tab:decomp_specs_const_caption}
\begin{center}
\begin{tabular}{|p{2cm}|p{0.8cm}|p{0.8cm}|p{1.2cm}|p{1.2cm}|}
\hline
\textbf{Target Gate} & \textbf{No. of Qubits} & \textbf{No. of Input Gates}  & \textbf{Maximum Allowed Depth} & \textbf{Optimal Depth}
\\ \hline
Reverse-CNOT & 2 & 4 & 5 & 3 \\ \hline
Magic \cite{Vatan_optimal} & 2 & 7 & 5 & 3 \\ \hline
Toffoli & 3 & 9 & 5 & 5 \\ \hline
CNOT$_{13}$ & 3 & 6 & 8 & 8 \\ \hline
Controlled-V & 2 & 9 & 7 & 6 \\ \hline
QFT & 2 & 9 & 8 & 6 \\ \hline
iSwap & 2 & 9 & 10 & 6 \\ \hline
Grover Diffusion & 2 & 12 & 6 & 6 \\ \hline
CNOT$_{41}$ & 4 & 6 & 10 & 10 \\ \hline
Fredkin \cite{smolin1996five} & 3 & 11 & 7 & 7 \\ \hline
\end{tabular}
\label{tab:gate_specs_const}
\end{center}
\end{table}

\begin{table}[ht]
\caption{Circuit decomposition specifications with angle parametrized gates.}
\label{tab:decomp_specs_param_caption}
\begin{center}
\begin{tabular}{|p{2cm}|p{0.8cm}|p{0.8cm}|p{1.2cm}|p{1.2cm}|}
\hline
\textbf{Target Gate} & \textbf{No. of Qubits} & \textbf{No. of Input Gates}  & \textbf{Maximum Allowed Depth} & \textbf{Optimal Depth}
\\ \hline
Hadamard & 2 & 9 & 3 & 1 \\ \hline
S (phase gate) & 2 & 32 & 3 & 1 \\ \hline
Controlled-Z & 2 & 72 & 4 & 3 \\ \hline
Magic & 2 & 73 & 4 & 2 \\ \hline
Controlled-H & 2 & 32 & 5 & 5 \\ \hline
W & 2 & 14 & 6 & 5 \\ \hline
Margolus & 3 & 12 & 7 & 7 \\ \hline 

\end{tabular}
\label{tab:gate_specs_param}
\end{center}
\end{table}

\subsection{Model and Solver Specifications}
In this paper, the two QCOpt models we compare are the MILP and the continuous NLP optimization models. Both of these models were solved to global optimality, and the NLP model was also solved to local optimality. All MILP models, implemented within QCOpt, were solved using Gurobi v9.5.2 \cite{gurobi} as the underlying solver with the solver's default parameters left unchanged. This solver combination is referred to as ``MILP Gurobi" in the remainder of this paper. 

The Gurobi solver also had recent developments in supporting global solutions via spatial branch-and-bound algorithms for non-convex bilinear problems \cite{Gurobi_breakthrough}. Thus, all the NLP relaxation models were solved to global optimality with Gurobi's ``Nonconvex'' option enabled, and with the solver's other default parameters left unchanged. This solver combination is referred to as ``NLP Gurobi" in the remainder of this paper. 

Finally, all the NLP relaxation models were also solved to local optimality using the Artelys Knitro solver v13.1, which is regarded as one of the most advanced and efficient solvers for large-scale non-linear optimization \cite{byrd2006k}. All of the Knitro parameters were left at their default values except for two: (1) we enabled Knitro's multi-warmstart strategy (via the `\texttt{mip\_multistart}' parameter) which was found to be important in facilitating the solver's convergence to a feasible solution, and (2) though in most cases, we had Knitro select the initialization strategy for NLP models, we found that for a few decompositions, selecting a linear relaxation initialization (via the `\texttt{ncvx\_qcqp\_init}' parameter) seemed to facilitate convergence to a feasible solution \cite{Knitro_options}. 
This model-solver combination is referred to as ``NLP Knitro'' in the remainder of the paper. 

\subsection{Numerical experiment specifications}

We make a few important notes before we discuss results in the next section. 
\begin{itemize}
    \item Both the MILP and NLP models solved the same exact input parameters for the decomposition as mentioned in Tables \ref{tab:decomp_specs_const_caption} and \ref{tab:decomp_specs_param_caption}. These are implemented in QCOpt's built-in library of standard target gate decompositions. 
    \item Both the MILP and NLP models included the symmetry-breaking valid constraints, as discussed in section \ref{subsec:symmetry}. Note that these valid constraints can vary based on the the input gates. 
    \item For each decomposition problem, all three model-solver combinations (MILP Gurobi, NLP Gurobi and NLP Knitro) produced identical circuits (up to the circuit symmetry) with the same optimal depth (listed in Tables \ref{tab:gate_specs_const} and \ref{tab:gate_specs_param}). 
    \item Even when the NLP models were solved only to local optimality (using the Knitro solver), we observed that the solutions obtained from the NLP models were in fact globally optimal. This observation was possible because the obtained circuits had the same depth as that of the solutions produced by globally solving the MILP model using the Gurobi solver.
    \item We were able to achieve convergence to integer solutions using NLP models for all the standard gate decompositions presented in this paper. This required an implementation of a ``multi start'' strategy, where multiple local searches were performed from randomly generated starting points. For this purpose, we randomly sampled from a uniform distribution of values from the respective variables' domains, that is,  $\mathrm{G}_d$ and $\widehat{\mathrm{G}}_d$ were sampled from $\mathcal{U}(-1,1)$. We also chose random binary assignments for the $z_{k,d}$ variables without any feasibility bias for these samples, except that every sample had to satisfy the constraint \eqref{eq:single_gate}.
    \item We chose to run `100' multi-starts for only the  NLP-Knitro model-solver combination, and chose the feasible solution with the minimum run time as shown in tables in section \ref{subsec:results}. Although we decided to capture the best run time for every gate, we observed that the variances in these run times across multiple starts, for the ones which converged to a feasible solution, were fairly low values.
    
\end{itemize}

\subsection{Performance of MILP vs. NLP models}
\label{subsec:results}
As Tables \ref{tab:run_times_const} and \ref{tab:run_times_param} illustrate, it is very clear that the `NLP Knitro' combination hands-down beats the `MILP Gurobi' and `NLP Gurobi' combinations, across all the standard gate decompositions with up to four qubit circuits. Note that the MILP-based linearization models implemented in QCOpt are efficient, and that Gurobi has implemented one of the best commercial MILP solvers. However, it was quite interesting to note that the NLP relaxation of the MINLP model (NLP model) was incredibly effective in solving the optimal quantum circuit design problem. Moreover, the reduction in run times did not imply any reduction in the quality of the feasible circuits. Instead, all the converged feasible circuits for the NLP model were indeed globally optimal when compared with respect to the MILP model's gate decompositions. 

\begin{table}[ht]
\caption{Circuit decomposition solver runtimes (in seconds) with Clifford+$\mathrm{T}$ gates. Here `NS' represents no solution. 
\label{runtimes_table}}
\begin{center}
\begin{tabular}{|p{2cm}|p{0.8cm}|p{0.8cm}|p{1.0cm}|p{0.8cm}|}
\hline
\textbf{Target Gate} & \textbf{MILP Gurobi} & \textbf{NLP Gurobi} & \textbf{NLP Knitro} & \textbf{Speed up}
\\ \hline
Reverse-CNOT & 0.02 & 1.03 & 0.02 & 1.0x \\ \hline
Magic & 0.06 & NS & 0.37 & 0.2x \\ \hline
Toffoli & 8.17 & NS & 3.19 & 2.6x \\ \hline
CNOT$_{13}$ & 11.99 & NS & 1.07 & 11.2x \\ \hline
Controlled-V & 26.0 & NS & 0.57 & 45.6x \\ \hline
QFT & 77.52 & NS & 3.43 & 22.6x \\ \hline
iSwap & 11.67 & NS & 5.93 & 2.0x \\ \hline
Grover Diffusion & 8.33 & NS & 0.94 & 8.9x \\ \hline
CNOT$_{41}$ & 30.59 & NS$^{\mathrm{a}}$ & 2.05 & 14.9x \\ \hline
Fredkin & 22.75 & NS$^{\mathrm{a}}$ & 4.98 & 4.6x \\ \hline \hline
\textbf{Average} & \textbf{19.7} & & \textbf{2.2} & \textbf{11.3x} \\ \hline 
\multicolumn{5}{l}{$^{\mathrm{a}}$The model had no solutions upon reaching a 300-second time limit.}
\end{tabular}
\label{tab:run_times_const}
\end{center}
\end{table}

\begin{table}[ht]
\caption{Circuit decomposition solver runtimes (in seconds) with angle parametrized gates.  Here `NS' represents no solution. 
\label{runtimes_table}}
\begin{center}
\begin{tabular}{|p{2cm}|p{0.8cm}|p{0.8cm}|p{1.0cm}|p{0.8cm}|}
\hline
\textbf{Target Gate} & \textbf{MILP Gurobi} & \textbf{NLP Gurobi} & \textbf{NLP Knitro} & \textbf{Speed up}
\\ \hline
Hadamard & 0.03 & 0.47 & 0.03 & 1.0x \\ \hline
S (phase gate) & 0.3 & 61.82 & 0.98 & 0.3x \\ \hline
Controlled-Z & 8.33 & NS$^{\mathrm{a}}$ & 6.93 & 1.2x \\ \hline
Magic & 26.7 & NS & 7.2 & 3.7x \\ \hline
Controlled-H & 3.23 & NS & 0.69 & 4.7x \\ \hline
W & 19.9 & NS$^{\mathrm{a}}$ & 1.72 &  11.6x \\ \hline
Margolus & 18.7 & NS$^{\mathrm{a}}$ & 2.13 &  8.8x \\ \hline \hline
\textbf{Average} & \textbf{11.0} & & \textbf{2.8} & \textbf{4.5x} \\ \hline
\multicolumn{5}{l}{$^{\mathrm{a}}$The model had no solutions upon reaching a 300-second time limit.}
\end{tabular}
\label{tab:run_times_param}
\end{center}
\end{table}

By contrast, the non-convex NLP model with Gurobi (`NLP Gurobi') did not perform well, even on smaller qubit circuits. We suspect this was primarily due to the fact that Gurobi applies global optimization techniques, which could lead to slower convergence.  

We shall now dive into the details of Tables \ref{tab:run_times_const} and \ref{tab:run_times_param}. On average, the run times of `NLP Knitro' performed \textit{11.3x} times faster than `MILP Gurobi' for input gates with constant entries (Clifford+$\mathrm{T}$ gates). Further, on average, the run times of `NLP Knitro' performed \textit{4.5x} times faster than `MILP Gurobi' for input gates parametrized with discretized angles. This lesser average improvement for the parametrized gates could be due to a substantial increase in the total number of input gates due to the parametrization options, as can be seen in Table \ref{tab:decomp_specs_param_caption}. We also suspect that the amount of symmetry created due to the parametrized rotation gates could be very high due to multiple same-axis rotations. Since all the symmetries may not be captured by the implemented symmetry-breaking constraints (section \ref{subsec:symmetry}) in QCOpt, that could also lead to an increase in the Gurobi and Knitro solvers' run times. 

It is also apparent that the capability of the `NLP Knitro' model-solver combination becomes clearer as the difficulty of the circuit decomposition increases. In particular, for more difficult target gate decompositions like `Quantum Fourier Transform (QFT)', `Controlled-V', `CNOT$_{41}$', `Fredkin' and `Margolus', `NLP Knitro' performed very well, with its biggest improvement being a \textit{45.6x} times faster convergence (with the `Controlled-V' gate) than the `MILP Gurobi' combination.

In summary, these intriguing results raise the possibility that the NLP relaxation model of the original MINLP model--solved by a local solver like Knitro--may be a very effective model-solver combination for the QCOpt package.
With further enhancements to QCOpt, as well as a deeper understanding of the behavior of the proposed NLP models and NLP solvers such as Knitro, we may be able to further push the boundaries of the QCOpt package to achieve the overarching goal of applying circuit decompositions to beyond Near-term Intermediate Scale Quantum (NISQ) - type devices. 

\section{Conclusions and Future Work}
\label{sec:conclusions}

In this paper, we have not only introduced multiple enhancements to QCOpt, but have also provided opportunities for exploring the behavior of state-of-the-art optimization-based solvers for non-linear programming problem models. Based on the MINLP formulation for the optimal quantum circuit design problem, we proposed a continuous non-linear programming (NLP) model by relaxing the integer variables. This NLP model allowed us to utilize efficient, large-scale barrier-based optimization methods in solvers like Knitro. Furthermore, although we solved these models to local optimality without integer requirements, we observed numerically  that all the solutions were indeed integral and also globally optimal, when compared with the implementation of QCOpt's MILP model. More importantly, the NLP model, solved by the Knitro solver, provided up to an \textit{11.3x} run time speed-up, on average, in comparison with the MILP model solved using the Gurobi solver.  

The avenues for future research are multitude: \textit{First}, we would like to obtain better theoretical insights into the NLP relaxation models, and to prove the property of obtaining integral solutions with them, if such a property exists. \textit{Second}, we observed that the convergence of a local solver to a feasible solution was very sensitive to the multi-start point which we provided based on a random sampling approach. We plan to investigate this further, with the goal of obtaining clusters of these starting points which could lead to a better convergence of local solvers. \textit{Third}, local solvers like Knitro offer a multitude of user parameters to choose from, and thus, to take full advantage of the solver's problem-specific capabilities would require a thorough investigation of its own. \textit{Fourth}, the performance of local solvers can also be quite sensitive to the type of NLP models we derive. Hence, we plan to explore multiple equivalent variants of these NLP models in order to find which behaves the best from the perspective of solver convergence. \textit{Finally}, we plan to explore machine learning-based approaches to potentially learn a map from the set of input and target gates to a high-quality starting point which could lead to quicker convergence to near-global optimal solutions within the NLP solver, akin to approaches in \cite{cengil2022learning}. 

\section{Acknowledgements}
This work was supported by Los Alamos National Laboratory's ``Laboratory Directed Research and Development (LDRD)'' program under projects ``20210114ER: Accelerating Combinatorial Optimization with Noisy Analog Hardware'' and ``20230091ER: Learning to Accelerate Global Solutions for Non-convex Optimization''.

\bibliographystyle{IEEEtran}
\bibliography{references.bib}

\end{document}